# A Numerical Analysis of a Micro- scale Piezoelectric Cantilever Beam: the Effect of Dimension Parameters on the Eigen Frequency

Tejas S. Fanse[1]

[1](Department of Mechanical Engineering, Texas A & M University, Kingsville, TX, USA)

**ABSTRACT:** *Eigen frequency is one of the most important system responses to be considered while designing a micro-scale piezoelectric cantilever beam. This paper investigates and analyzes the effect of dimension parameters of a micro-scale piezoelectric cantilever beam on its eigen frequency. The beam is assumed to be made of silicon. Structural mechanics based finite element analysis is carried out in the environment of COMSOL Multiphysics software for the study. It is seen that length and thickness have continuous effect on the eigen frequency; whereas, the effect of width is discontinuous.*

---

---

## I.    INTRODUCTION

Application of micro-electro-mechanical system based piezoelectric actuation is increasing day by day in many applications— from compact electronic devices to industrial machineries. Designing and modeling of the micro-scale cantilever beam is, in many of the cases, a prerequisite for the optimized fabrication of micro-electro-mechanical system based piezoelectric actuators. Finite element method is a very reliable, fast and accurate one for this purpose.

There have been many researches related to design and modeling of different piezoelectric beam configurations— most of which were conducted with finite element method. Extension and shear piezoelectric actuation have been investigated with finite element model [1]. Piezoelectric active structures have been investigated with finite element modeling [2]. Exact solutions have been proposed for the electromechanical response of piezoelectric actuators [3]. Theoretical modeling was proposed for piezoelectric multi- morph [4]. Frequency response was analyzed for metallic cantilever beam coupled with piezoelectric transducer [5]. Tow-dimensional model has been proposed for thick piezoelectric actuator [6]. Adaptive control of piezoelectric cantilever beam has been researched [7]. Film piezoelectric cantilevers have been used for energy harvesting [8]. Piezoelectric cantilever composite plates have been investigated [9]. Multi-layered piezoelectric cantilever beams have been analyzed [10]. Eigen frequency analysis was conducted for sound-structure interaction in rectangular enclosure [11]. Vibration energy harvesting for low frequency using micro-machined PZT cantilever has been researched [12]. Complex electrode configuration of piezoelectric elements has been modeled with finite element method [13]. Deformation characteristics of circular beam with asymmetric piezoelectric actuators have been analyzed [14]. Effect of length has been analyzed for uni-morph cantilevers [15]. Piezoelectric vibrating generator of cantilever has been modeled and analyzed [16]. Beams with piezoelectric sensors and actuators have been analyzed [17]. Effect of beam length on eigen frequency has been analyzed for piezo-ceramic materials [18].

It should be noted that eigen frequency is a very important system response to be considered for designing a piezoelectric cantilever beam. It is because the output energy becomes the maximum only if the ambient frequency becomes equal to the eigen frequency of the energy harvester, i.e., eigen frequency of the cantilever beam. This paper analyzes the effect of dimension parameters of a silicon based piezoelectric cantilever beam on its eigen-frequency. This analysis will help the designers to design a piezoelectric cantilever beam more easily and accurately in accordance with its dimension and eigen frequency.

## II.    COMPUTATIONAL MODELING AND NUMERICAL ANALYSIS

The computational modeling and the numerical analysis for this study were carried out in the environment of COMSOL Multiphysics software. Fig. 1 shows the three-dimensional view of a random phase geometric model of the piezoelectric cantilever beam to be analyzed. The length, the width and the height of this beam shown in Fig.1 is 300 μm, 40 μm and 7 μm respectively. The material of the beam is considered to be silicon. The density, the young's modulus and the Poison's ratio for silicon were considered to be 2329 kg/m$^3$, 17×10$^{10}$ Pa and 0.28. Governing equations considered for this analysis were as follow.

Eigen frequency is described by the equation,

—qm$^2$u— A. a = Fv      (1)

Here, q is the distributed free charge, u is the displacement vector, a is the Cauchy stress tensor, F describes the body force vector and assuming Z to be frequency,

—im = h            (2)

Again, for fixed constraint,

u = O      (3)

Cauchy stress tensor,

a = a$_O$ + c: (s — s$_O$ — s$_{inSet}$)           (4)

Strain displacement equation for infinitesimal strain tensor is given by,

s = 0.5[Au + (Au)$^T$]        (5)

These equations were used for solution by the structural element analysis module that was used for the analysis. As, it was a cantilever beam, one end surface of the beam was considered to be fixed, while all other surfaces of the beam were considered to be free. The meshing was selected to be physics-based meshing of extremely fine quality.

## III.   FIGURES AND TABLES

For the simulation, two of the dimensions were assumed fixed, whereas, the other one was varied as suggested in Table I.

**TABLE I.**PARAMETERS FOR CONSIDERATION

| Varying parameter | Length (L) μm | Width (W) μm | Height (H) μm |
|---|---|---|---|
| Length (L) | 100~1000 | 40 | 7 |
| Width (W) | 300 | 10~100 | 7 |
| Height (H) | 300 | 40 | 1~10 |

It is seen that for fixed condition— length, width and height are assumed to be 300 μm, 40 μm and 7 μm respectively. The first six eigen frequencies are calculated numerically for each of the situation. Fig. 2 to Fig. 5 shows the mode shape plot of the micro-scale cantilever beam for its eigen frequency. The dimension of the beam considered here is 100 × 40 × 7 (L × W × H).

Fig. 2 shows mode shape plot for the first eigen frequency mode, while Fig. 3 shows it for the second eigen frequency mode. Fig. 4 shows the third eigen frequency mode while Fig. 5 shows the fourth eigen frequency mode for the same dimension. The vertical color bar denotes the displacement of the structure at each point. Just like these first four modes of frequencies of this fixed dimension, all the plots for the six eigen frequencies can be determined for the variable dimensions considered for the whole analysis.

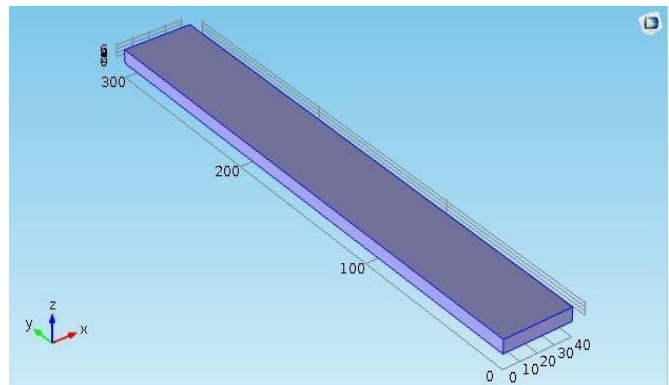

Fig. 1. Three-dimensional view of the geometric model of the micro-scale piezoelectric cantilever beam

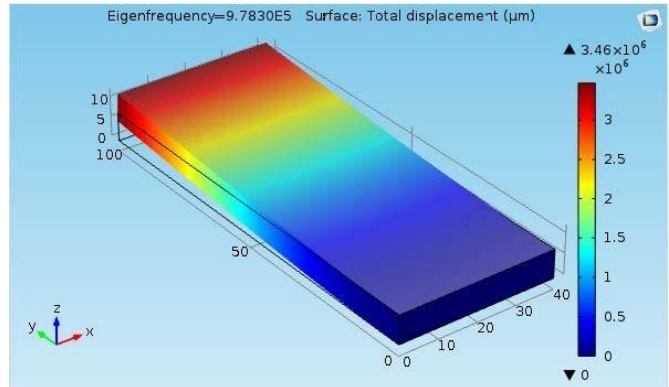

Fig. 2. First frequency mode for the micro-scale cantilever
beam with the dimension of $100 \times 40 \times 7$ (L × W × H)

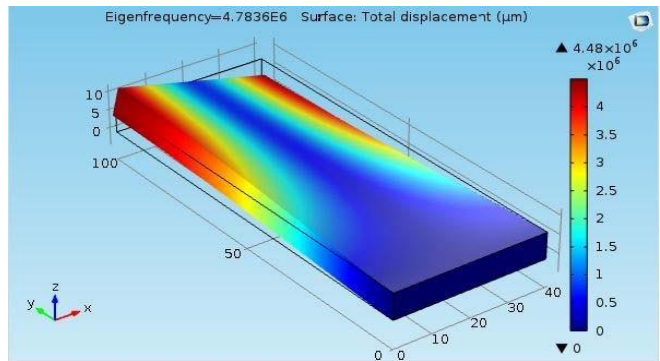

Fig. 3. Second frequency mode for the micro-scale cantilever
beam with the dimension of $100 \times 40 \times 7$ (L × W × H)

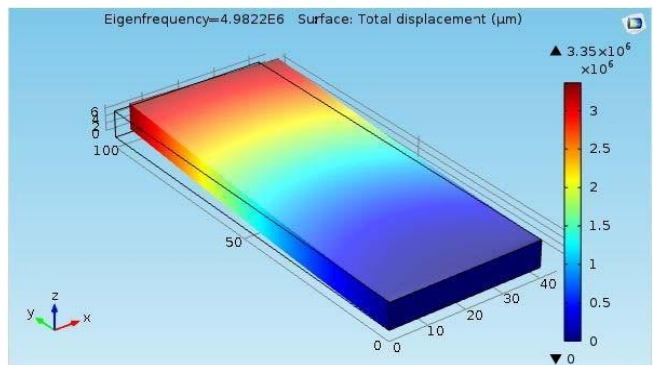

Fig. 4. Third frequency mode for the micro-scale cantilever
beam with the dimension of $100 \times 40 \times 7$ (L × W × H)

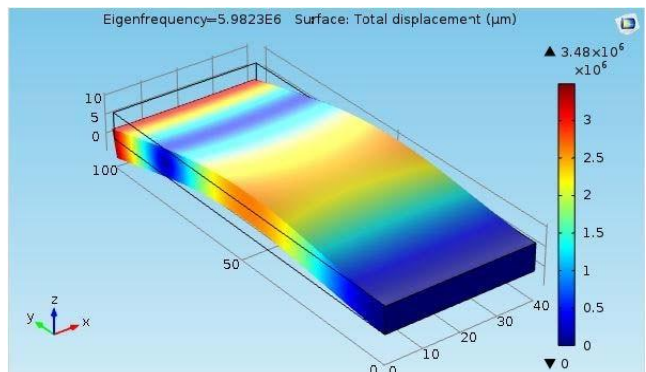

Fig. 5. Fourth frequency mode for the micro-scale cantilever
beam with the dimension of $100 \times 40 \times 7$ (L × W × H)

Based on the numerical analysis, results are plotted in the graphs. Fig. 6 shows the effect of length of the micro-scale cantilever beam on eigen frequency of the beam. It is seen that for a specific mode of eigen frequency, if the length is increased, the eigen frequency decreases. It is also seen that, for lower value of length, the difference of one mode frequency to another is quite high, while, increase of length of the beam leads to convergence of the all six modes of frequency.

For the current study, for the minimum of length, i.e., for the length of 100 μm, the first mode frequency is $9.78 \times 10^5$ Hz and the sixth mode frequency is $1.62 \times 10^7$ Hz, while, for the maximum length, i.e., for the length of 1000 μm, the first mode frequency is $9.68 \times 10^3$ Hz and sixth mode frequency is $3.43 \times 10^5$ Hz.

Fig. 7 shows the effect of width of the micro-scale cantilever beam on eigen frequency of the beam. It is seen that the first mode frequency and the fifth mode frequency is almost constant for any considerable width of the beam. Other mode frequencies are not necessarily constant, but the overall effect of width of the beam on its eigen frequency can be considered to be discontinuous.

Fig. 8 shows the effect of height of the micro-scale cantilever beam on eigen frequency of the beam.

It is seen that for a specific mode of eigen frequency, if the height is increased, the eigen frequency increases. It is also seen that, for lower value of height, the difference of one mode frequency to another is quite small, while, increase of length of the beam leads to a long difference of the mode frequencies.

For the current study, for the minimum of height, i.e., for the height of 1 μm, the first mode frequency is $1.54 \times 104$ Hz and the sixth mode frequency is $6.06 \times 105$ Hz, while, for the maximum height, i.e., for the height of 10 μm, the first mode frequency is $1.54 \times 105$ Hz and sixth mode frequency is $3.53 \times 106$ Hz.

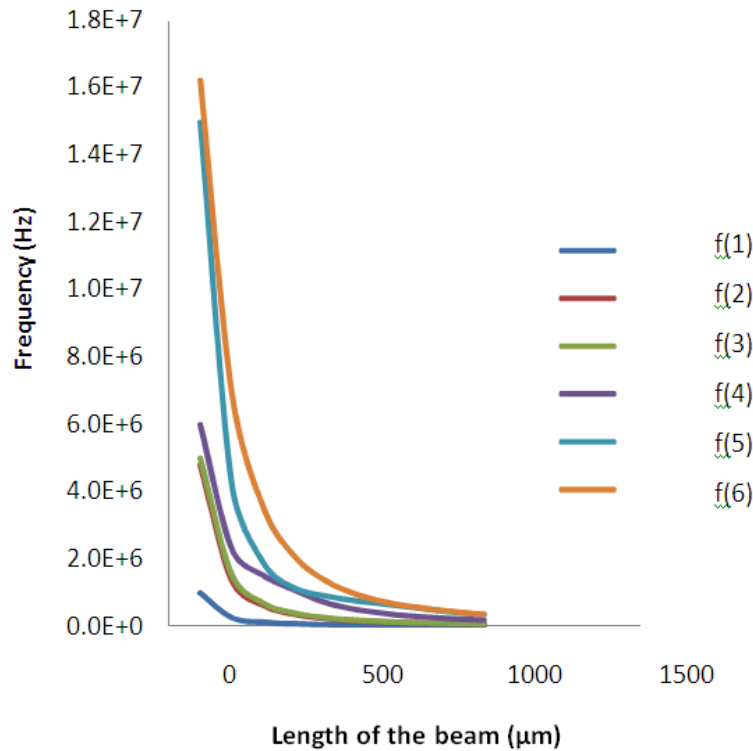

Fig. 6. Effect of length on the eigen frequency of the micro-scale cantilever beam

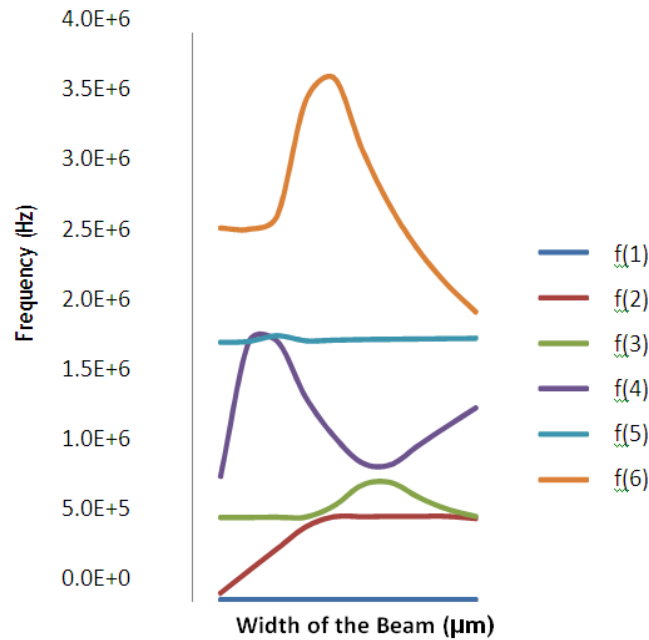

**Fig. 7.** Effect of width on the eigen frequency of the micro-scale cantilever beam

## IV.    CONCLUSION

Eigen frequency is a very important system response to be considered for designing piezoelectric actuators. In the current study, the effect of length, height and width: all three of the dimension parameters of a micro-scale piezoelectric cantilever beam have been varied for investigating their effect on the eigen frequency of that beam in the environment of COMSOL Multiphysics software. It is seen that the effect of length and height of the beam is continuous on the eigen frequency, but the effect of width is discontinuous on the eigen frequency. This study can be further used by the researchers for designing micro-scale piezoelectric cantilever beam.